\documentclass[12pt,aps,pra,reprint,showpacs]{revtex4-1}
\usepackage{bm,color,amsmath,amssymb,mathrsfs,latexsym,graphicx,psfrag}

\def\be{\begin{equation}}
\def\ee{\end{equation}}
\def\ba{\begin{eqnarray}}
\def\ea{\end{eqnarray}}

\begin{document}

\title{Spin-triplet superconductivity in a weak-coupling Hubbard model for the quasi-one-dimensional compound Li$_{0.9}$Mo$_6$O$_{17}$}
\author{Weejee Cho${}^{1}$}
\author{Christian Platt${}^{1}$}
\author{Ross H. McKenzie${}^{2}$}
\author{Srinivas Raghu${}^{1,3}$}
\affiliation{${}^1$ Department of Physics, Stanford University, Stanford, CA 94305, USA\\
${}^2$ School of Mathematics and Physics, University of Queensland, Brisbane, 4072 Queensland, Australia\\
${}^3$ Stanford Institute for Materials and Energy Sciences, SLAC National Accelerator Laboratory, Menlo Park, California 94025, USA
}

\date{\today}
\begin{abstract}
The purple bronze Li$_{0.9}$Mo$_6$O$_{17}$ is of interest due to its quasi-one-dimensional electronic structure and the possible Luttinger liquid behavior resulting from it. For sufficiently low temperatures, it is a superconductor with a pairing symmetry that is still to be determined.  To shed light on this issue, we analyze a minimal Hubbard model for this material involving four molybdenum orbitals per unit cell near quarter filling, using asymptotically exact perturbative renormalization group methods. We find that spin triplet odd-parity superconductivity is the dominant instability. Approximate nesting properties of the two quasi-one-dimensional Fermi surfaces enhance certain second-order processes, which play crucial roles in determining the structure of the pairing gap.  Notably, we find that the gap has more sign changes than required by the point-group symmetry.
\end{abstract}

\maketitle

\section{Introduction}
Understanding unconventional superconductors, characterized by
a sign-changing order parameter, remains a challenge \cite{Maiti2013,norman2011}.
Most known unconventional superconductors (cuprates, organic charge transfer salts, iron pnictides) involve pairing in the spin-singlet channel. The leading candidates for spin-triplet pairing are strontium ruthenate \cite{Maeno2012}
and the heavy-fermion compound UPt$_3$. In general, it remains unclear how the pairing symmetry is determined by the interplay of strong electron correlations, low dimensionality, spin fluctuations \cite{scalapino2012}, multiple bands \cite{Raghu2010}, Hund's rule coupling, band structure effects \cite{cho2013}, and spin correlations in proximate Mott-insulating states \cite{powell2007}.

Li$_{0.9}$Mo$_6$O$_{17}$, known more commonly as a ``purple bronze,'' is a layered transition metal oxide that exhibits several exotic properties arising from the effective low dimensionality and strong electron correlations.  At high temperatures, the dynamics are primarily one-dimensional and the system exhibits some
properties that have been interpreted in a Luttinger liquid framework \cite{dudy2013,podlich,chudzinski2012}.
At lower temperatures the metallic phase is difficult to characterize, possibly due to incipient
charge density wave order \cite{mckenzie2012prb,merino2014}. Still, for magnetic fields perpendicular to the easy transport axis (the crystalline $b$ axis), the magnetoresistance grows without bound as a function of the field strength, as consistent with an open Fermi surface \cite{chen10}. There are several unconventional aspects of the superconducting state  that are worth noting.  First, the transition temperature $T_c$ decreases with increasing disorder
(residual resistivity) \cite{matsuda1986}, a signature of non-$s$-wave pairing (for a review, see Ref. \onlinecite{powell2004}).
Second, the upper critical field exhibits a great deal of anisotropy, consistent with a quasi-one-dimensional system.  In particular, for a magnetic field applied along the $b$ axis, the upper critical field exceeds the Chandrasekhar-Clogston paramagnetic limit \cite{mercure2012}, suggesting that the superconductivity may be spin-triplet in character. However, this requires ruling
out alternative explanations such as
the effects of spin-orbit coupling and quasi-one-dimensional fluctuations \cite{fuseya2012}.

A well-studied example of unconventional pairing in quasi-one-dimensional systems is the organic Bechgaard salts (TMTSF)$_2$X.  In these materials, there is evidence in favor of nodal, spin-singlet pairing \cite{shinagawa07prl}.  However, in the presence of a strong  in-plane magnetic field applied along the chains, superconductivity in these systems also appears to exceed the Chandrasekhar-Clogston limit.  If a modest magnetic field can induce a transition of this sort, it is quite natural to infer that there is a near degeneracy of pairing in the spin-singlet and spin-triplet channels.  Indeed, weak-coupling considerations based on Fermi surface nesting do indicate such a near degeneracy in these systems \cite{bourbonnais06prb,cho2013}.
Whether the pairing is singlet or triplet remains a subject of debate \cite{Zhang2007}.
Here we explore the role of Fermi surface nesting,
multiband effects, and low dimensionality for
unconventional superconductivity in the context of  Li$_{0.9}$Mo$_6$O$_{17}$.

Despite evidence of strong correlation effects in purple bronze, we take here a weak-coupling perspective.  The advantage of such an approach is that the solutions obtained are well controlled, and robust trends suggesting a certain pairing symmetry can easily be identified in this limit.  The qualitative information obtained this way may be relevant to the experimentally relevant intermediate-coupling regime. Indeed, for the cuprates it has been found that the weak-coupling predictions of $d$-wave superconductivity are consistent
with spin-fluctuation approaches and strong-coupling approaches
such as RVB theory \cite{scalapino2012}. For simplicity, we consider a four-band model Hamiltonian with on-site repulsive electron interactions.

Our principal findings may be summarized as follows: Unlike the case of Bechgaard salts, the near degeneracy between the dominant triplet and the dominant singlet solutions is absent; the trivial pairing channel is significantly penalized, as the bare on-site repulsion takes effect in multiple sublattices.
Significantly, the leading
pairing instability  is in the odd-parity channel; there is a sign change in the gap function (not required by symmetry) across the two nearly degenerate Fermi surfaces, and additional accidental nodes develop in the inner surface when the parameter inducing the splitting of the two Fermi surfaces as well as their warping is sufficiently large.

\begin{figure}[t]
\includegraphics[scale=0.75]{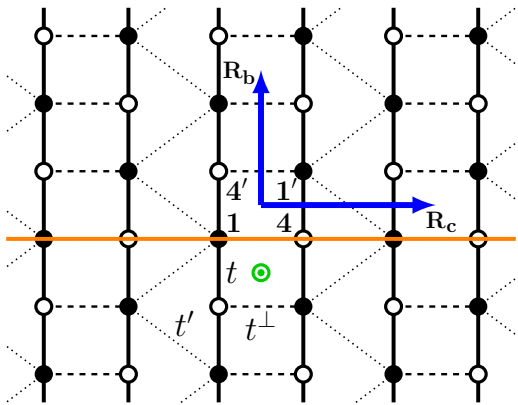}\\
\caption{Tight-binding lattice model. Each circle corresponds to
a single Mo atom and there are four atoms per unit cell. Filled and empty circles denote two types of crystallographically inequivalent Mo atoms, i.e., Mo(1) and Mo(4), within
a layer of Li$_{0.9}$Mo$_6$O$_{17}$.
The horizontal orange line defines a plane of mirror symmetry, and the green dot an axis of $C_2$-rotation symmetry. (The reflection is only an approximate symmetry in the real material, which is slightly monoclinic \cite{onoda1987jssc}.)
}
\label{ladder}
\end{figure}

\section{The model}
In Li$_{0.9}$Mo$_6$O$_{17}$, the low-energy electronic degrees of freedom reside on ladders built from $d_{xy}$ orbitals of Mo atoms \cite{whangbo1988jacs,Popovic:2006,jarlborg2012}, which run along the crystallographic $b$ axis. Based on this observation, the electronic structure was recently described by a four-band tight-binding model near quarter filling, i.e., $n_{\textmd{el}}=1.9$ out of 8 per unit cell, in Ref. \onlinecite{mckenzie2012prb}. (Similar, but slightly different
models have been presented in Refs. \onlinecite{chudzinski2012} and
\onlinecite{nuss2014}.) In this model, bonds within each chain are formed by the hopping amplitude $t$, whereas $t^{\perp}$ connects two nearest chains to make a ladder, and neighboring ladders are coupled by $t^{\prime}$ (Fig. \ref{ladder}). The resulting single-particle Hamiltonian can be represented in the momentum space as follows:
\begin{widetext}
\begin{equation}
h_{\tau\tau^{\prime}}(\textbf{k}) = -\left(\begin{array}{cccc}0&t^{\perp}&t^{\prime}e^{-ik_{c}}(1+e^{-ik_{b}})&t(1+e^{-ik_{b}})
\\t^{\perp}&0&t(1+e^{-ik_{b}})&0\\t^{\prime}e^{ik_{c}}(1+e^{ik_{b}})&t(1+e^{ik_{b}})&0&t^{\perp}\\t(1+e^{ik_{b}})&0&t^{\perp}&0\end{array}\right).
\end{equation}
\end{widetext}
Here, $\tau$ and $\tau^{\prime}$ are sublattice indices, and in the basis used to construct the above matrix, the order of sublattices is 1, 4, 1${}^\prime$, and 4${}^\prime$. (See Fig. \ref{ladder}.) We denote the $n^{\textmd{th}}$ eigenvalue of $h_{\tau\tau^{\prime}}(\textbf{k})$ as $\alpha_{\tau}(n,\textbf{k})$ and the corresponding band dispersion as $\epsilon(n,\textbf{k})$.

So long as $|t^{\perp}|, |t^{\prime}|\ll |t|$ and the filling fraction is near one-quarter, two nearly degenerate Fermi surfaces exist. They are open along the $c$ axis, and there is an approximate interband nesting betweeen them at $\textbf{Q} \equiv (0,\pi n_{\textmd{el}}/2) = (0,0.95\pi)$ (Fig. \ref{fs}). To identify a robust tendency that does not involve a fine-tuning of the tight-binding band structure, we introduce the warping parameter $\eta_{w}$, where $t^{\perp}=-0.048\eta_{w}t$ and $t^{\prime}=0.072\eta_{w}t$. As $\eta_{w}$ is increased, the splitting between the two Fermi surfaces and the warping of them become larger; at the same time, the nesting becomes less perfect. The band structure is reduced to that described in Ref. \onlinecite{mckenzie2012prb} at $\eta_{w}=1$.

We further add on-site repulsive interactions that lead to a Hubbard model with the following Hamiltonian:
\begin{equation*}
H = H_{0}+H_{\textmd{int}},
\end{equation*}
\begin{equation}
H_{0} = \sum_{\textbf{k}, \tau, \tau^{\prime}, \sigma} h_{\tau\tau^{\prime}}(\textbf{k})\, c_{\textbf{k}\tau\sigma}^{\dagger}c_{\textbf{k}\tau^{\prime}\sigma},
\end{equation}
\begin{equation}
\label{Hint}
H_{\textmd{int}} =\frac{U}{N}\!\!\sum_{\{\textbf{k}_{i}\},\tau} c_{\textbf{k}_{1}\tau\uparrow}^{\dagger}c_{\textbf{k}_{2}\tau\downarrow}^{\dagger} c_{\textbf{k}_{4}\tau\downarrow} c_{\textbf{k}_{3}\tau\uparrow},
\end{equation}
with the understanding that $\textbf{k}_{4} \equiv \textbf{k}_{1}+\textbf{k}_{2}-\textbf{k}_{3}$. Spin indices are denoted as $\sigma$ and $\sigma^{\prime}$, and number of unit cells as $N$.

The weak-coupling regime corresponds to the case that $|t|, |t^{\perp}|, |t^{\prime}|\gg U$.
In this limit, it is desirable to work in the band basis that makes $H_{0}$ diagonal as follows:
\begin{equation}
\label{bandbasis}
c_{\textbf{k}\tau\sigma} = \sum_{n} \alpha_{\tau}(n,\textbf{k}) \,c_{n \textbf{k}\sigma},
\end{equation}
and
\begin{equation}
H_{0}=\sum_{n,\textbf{k},\sigma} \epsilon(n,\textbf{k})\,c_{n\textbf{k}\sigma}^{\dagger}c_{n\textbf{k}\sigma}.
\end{equation}

\section{Perturbative renormalization group (RG) method}
We implement a renormalization group (RG) method \cite{raghu2010weakcoupling,Raghu2010,cho2013} to investigate superconducting instabilities {\'a} la Kohn and Luttinger \cite{kohn-65prl524}. The effective interaction at energy scales asymptotically close to the Fermi surface is given by the following expression:
\begin{widetext}
\begin{equation}
\label{pairingvertex}
\begin{split}
&\Gamma_{\uparrow\downarrow} (n,\hat{\textbf{k}};n^{\prime},\hat{\textbf{k}}^{\prime}) = \Gamma_{\uparrow\downarrow}^{(1)} (n,\hat{\textbf{k}};n^{\prime},\hat{\textbf{k}}^{\prime})
+\Gamma_{\uparrow\downarrow}^{(2)} (n,\hat{\textbf{k}};n^{\prime},\hat{\textbf{k}}^{\prime}),\\
&\Gamma_{\uparrow\downarrow}^{(1)} (n,\hat{\textbf{k}};n^{\prime},\hat{\textbf{k}}^{\prime}) \equiv U \sum_{\tau} |\alpha_{\tau}(n,\hat{\textbf{k}})|^{2} |\alpha_{\tau}(n^{\prime},\hat{\textbf{k}}^{\prime})|^{2},\\
&\Gamma_{\uparrow\downarrow}^{(2)} (n,\hat{\textbf{k}};n^{\prime},\hat{\textbf{k}}^{\prime}) \equiv - U^{2}\sum_{m,m^{\prime}} \int_{BZ} \frac{d^{2}\textbf{p}}{(2\pi)^{2}}\, \Bigg\{
\frac{f\big(\epsilon(m^{\prime},\textbf{p})-\mu\big)-f\big(\epsilon(m,\textbf{p}+\hat{\textbf{k}}+\hat{\textbf{k}}^{\prime})-\mu\big)}{\epsilon(m^{\prime},\textbf{p})-\epsilon(m,\textbf{p}+\hat{\textbf{k}}+\hat{\textbf{k}}^{\prime})}\\
&\qquad\qquad\qquad\qquad\qquad\qquad\bigg|\sum_{\tau}\alpha_{\tau}(n,\hat{\textbf{k}})^{\ast} \,\alpha_{\tau}(m,\textbf{p}+\hat{\textbf{k}}+\hat{\textbf{k}}^{\prime}) \, \alpha_{\tau}(m^{\prime},\textbf{p})^{\ast} \, \alpha_{\tau}(n^{\prime},\hat{-\textbf{k}}^{\prime}) \bigg|^{2}
\Bigg\}.
\end{split}
\end{equation}
\end{widetext}
Here, $f$ is the Fermi function at zero temperature, and $\mu$ is the chemical potential. It represents a process where two electrons of opposite spins scatter from $(n^{\prime},\pm\hat{\textbf{k}}^{\prime})$ to $(n,\pm\hat{\textbf{k}})$ without spin flip. It is unnecessary to consider the corresponding object for equal-spin electrons: due to spin-rotation invariance, Eq. (\ref{pairingvertex}) accounts for both the singlet and triplet channels.

Different pairing channels are identified by solving the following integral eigenequation defined along the Fermi surface:
\begin{equation}
\begin{split}
\label{integraleq}
\sum_{n^{\prime}}&\int\!\frac{ds(\hat{\textbf k}_{n^{\prime}}^{\prime})}{(2\pi)^{2}v(n^{\prime},\hat{\textbf k}_{n^{\prime}}^{\prime})}\Gamma_{\uparrow\downarrow}(n,\hat{\textbf k}_{n};n^{\prime},\hat{\textbf k}_{n^{\prime}}^{\prime})
\,\psi_{\alpha}(n^{\prime},\hat{\textbf k}_{n^{\prime}}^{\prime})\\
&=\lambda_{\alpha} \psi_{\alpha}(n,\hat{\textbf k}_{n}).
\end{split}
\end{equation}
Here, $\hat{\textbf k}_{n}$ denotes a point on the Fermi surface in the $n^{\textmd{th}}$ band, $ds(\hat{\textbf{k}}_{n})$ is an infinitesimal length element of this Fermi surface, and $v$ is the Fermi velocity. The pair wave function $\psi_{\alpha}$ is classified according to its parities under the transformations $k_{c} \rightarrow -k_{c}$ and $k_{b} \rightarrow -k_{b}$. These symmetry operations in the momentum space are derived from the symmetry in the real space shown in Fig. \ref{ladder}.

Negative eigenvalues of $\Gamma_{\uparrow\downarrow}$ indicate the sectors in which the effective interaction is attractive. These attractive interactions grow under further renormalization. The eigenfunction with the most negative eigenvalue (denoted as $\lambda_{0}$) is responsible for superconductivity; pairing instability sets in at the energy scale of $W e^{-1/|\lambda_{0}|}$, where $W\equiv 4t$ is the bandwidth. This energy scale is also identified with the transition temperature $T_{c}$, and the gap structure is inherited from $\psi_{0}(n,\hat{\textbf k}_{n})$.

Notice that in the asymptotic weak-coupling limit, the bare on-site repulsion $\Gamma_{\uparrow\downarrow}^{(1)}$ in Eq. (\ref{pairingvertex}) is infinitely larger than the other term $\Gamma_{\uparrow\downarrow}^{(2)}$. Hence, attraction can arise only if the pair wave function completely avoids the effect of $\Gamma_{\uparrow\downarrow}^{(1)}$, i.e., belongs to its null space:
\begin{equation}
\label{constraints}
\sum_{n}\int\!\frac{ds(\hat{\textbf k}_{n})}{(2\pi)^{2}v(n,\hat{\textbf k}_{n})} |\alpha_{\tau}(n,\hat{\textbf{k}}_{n})|^{2} \psi_{\alpha}(n,\hat{\textbf k}_{n}) = 0
\end{equation}
for all four possible values of the sublattice index $\tau$. In practice, one should solve Eq. (\ref{integraleq}) with $\Gamma_{\uparrow\downarrow}$ replaced by $\Gamma_{\uparrow\downarrow}^{(2)}$ while enforcing the above conditions.

\begin{figure}[b]
\centering
\includegraphics[scale=1]{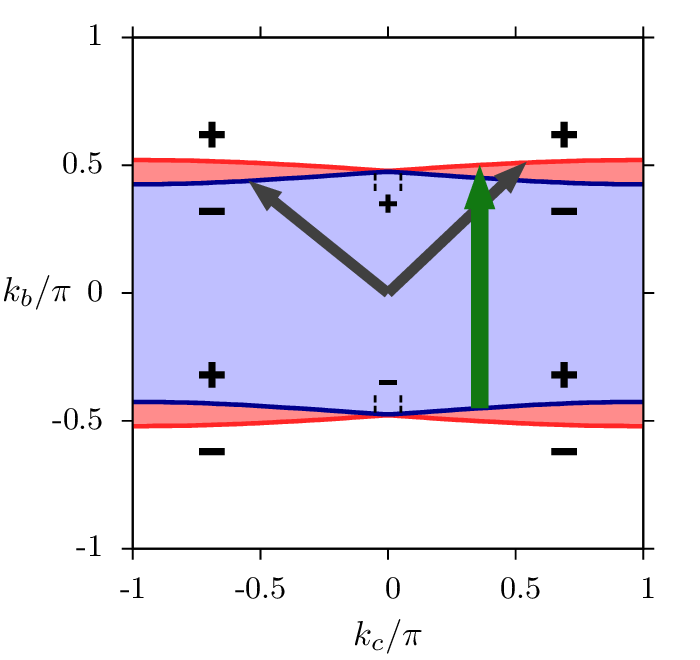}
\caption{The Fermi surface and the sign structure of the (triplet) gap function for $\eta_{w}=1$. The inner and outer curves correspond to Fermi surface sheets associated with distinct bands. The vertical arrow depicts the nesting vector $\textbf{Q}\equiv(0, 0.95\pi)$.
The oblique arrows denote an example of two points on the Fermi surface whose crystal momenta approximately sum to $\textbf{Q}$; the gap function changes sign between these points although not required by symmetry.}
\label{fs}
\end{figure}

It should be noted that Eq. (\ref{constraints}) is automatically satisfied by pair wave functions that are odd under $k_{c} \rightarrow -k_{c}$ or $k_{b} \rightarrow -k_{b}$. To see this, observe that $|\alpha_{\tau}(n,\hat{\textbf{k}}_{n})|^{2}$ is an even function of both $k_{c}$ and $k_{b}$. This fact is a result of the reflection (see Fig. \ref{ladder}) and time-reversal symmetries. When $\psi_{\alpha}(n,\hat{\textbf k}_{n})$ is an odd function of either $k_{c}$ or $k_{b}$, the contributions from different quadrants of the Brillouin zone cancel out in the integral in Eq. (\ref{constraints}). On the other hand, the four conditions given by Eq. (\ref{constraints}) lead to additional constraints on solutions that are even in both $k_{c}$ and $k_{b}$. However, due to the $C_{2}$-rotation symmetry, Eq. (\ref{constraints}) becomes identical for $\tau = 1, 1^{\prime}$ as well as for $\tau = 4, 4^{\prime}$ (Fig. \ref{ladder}), and the number of constraints is two as opposed to four.

In addition to constraints set by the bare repulsion, nesting properties of the Fermi surface play an important role in determining the gap structure. In our case, $\Gamma_{\uparrow\downarrow}^{(2)} (n,\hat{\textbf{k}};n^{\prime},\hat{\textbf{k}}^{\prime})$ in Eq. (\ref{pairingvertex}) is considerably enhanced when the momentum transfer $\hat{\textbf{k}}+\hat{\textbf{k}}^{\prime}$ is close to the approximate nesting vector $\textbf{Q} = (0.95\pi,0)$ (Fig. \ref{fs}). Moreover, from its definition, we see that $\Gamma_{\uparrow\downarrow}^{(2)}$ is always positive. Summing up, the nesting condition $\hat{\textbf{k}}+\hat{\textbf{k}}^{\prime} \approx \textbf{Q}$ leads to a large positive off-diagonal matrix element connecting $\hat{\textbf{k}}$ and $\hat{\textbf{k}}^{\prime}$. Therefore, one can expect that a pair wave function with a large negative eigenvalue tends to change sign between two points on the Fermi surface whose crystal momenta sum to $\textbf{Q}$.

Together with the repulsive Hubbard interaction, the nesting property of the Fermi surface shown in Fig. \ref{fs} causes spin density wave (SDW) fluctuations of wave vector $\textbf{Q}$. In view of this fact, the previous paragraph is about how SDW fluctuations enhance superconductivity and determine the gap structure. Notice that within our weak-coupling approach, such nonsuperconducting fluctuations are never strong enough to compete with superconductivity unless an extreme fine-tuning is made. That is, as long as the band dispersion is invariant under $\textbf{k}\,\to\,-\textbf{k}$, the perfect nesting in the particle-particle channel makes superconductivity dominant over other types of instabilities except at certain special points in the parameter space (e.g., Van Hove singularity, perfect nesting in the particle-hole channel). The one-dimensional limit ($\eta_{w}=0$) corresponds to such a special point, but we avoided it. Then, as we know that the ground state is superconducting, it is indeed a valid approach to investigate the form of pairing that leads to the strongest instability of the Fermi sea without considering nonsuperconducting instabilities. Correct description of the 1D physics at $\eta_{w}=0$ is beyond the scope of this study.

\section{Results of the weak-coupling analysis}
\begin{figure}[t]
\centering
\includegraphics[scale=0.85]{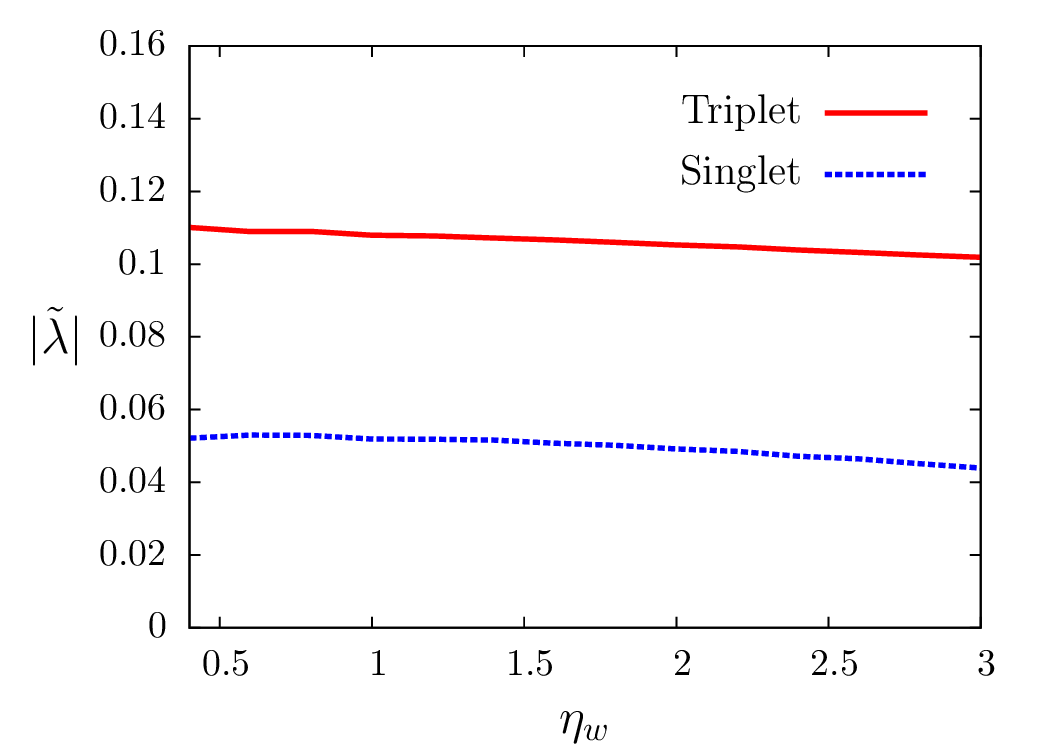}
\caption{Dominance of triplet pairing is insensitive to the extent of warping of the
Fermi surface.
The pairing strength in the triplet and singlet channels is shown as a function
of the parameter $\eta_w$ that determines warping of the Fermi surface.
Reference \onlinecite{mckenzie2012prb} estimates that $\eta_w \simeq 1$.
}
\label{pairing}
\end{figure}

\begin{figure}[t]
\centering
\includegraphics[scale=1]{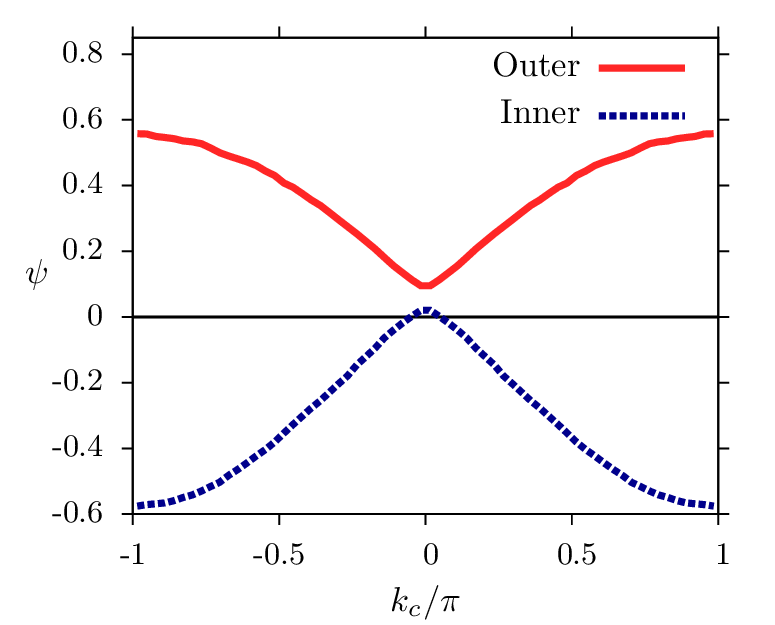}
\caption{The pair wave function for $\eta_{w}=1$ on the inner and outer Fermi surfaces in the upper half of the Brillouin zone.}
\label{pairwave}
\end{figure}

Figure \ref{pairing} shows the dominant pairing strength in the triplet and singlet channels as functions of the warping parameter $\eta_{w}$. (We have plotted the dimensionless quantity $\tilde{\lambda} \equiv \lambda \frac{W^{2}}{U^{2}}$.) To obtain this result, we numerically solved Eq. (\ref{integraleq}) by discretizing each of the two Fermi surfaces into 128 segments and evaluated $\Gamma_{\uparrow\downarrow}^{(2)}$ in Eq. (\ref{pairingvertex}) by discretizing the Brillouin zone into a 2048$\times$2048 grid. \footnote{Notice that in the expression for $\Gamma_{\uparrow\downarrow}^{(2)}$, the support of the integrand is a tiny fraction of the Brillouin zone if $m=m^{\prime}$ and the momentum transfer is small. In this case, we transformed the expression into an integral over the Fermi surface using the following relation:
\begin{equation*}
\begin{split}
&\int_{BZ} \frac{d^{2}\textbf{p}}{(2\pi)^{2}} \frac{f\big(\varepsilon(\textbf{p})\big)-f\big(\varepsilon(\textbf{p}+\textbf{q})\big)}{\varepsilon(\textbf{p})-\varepsilon(\textbf{p}+\textbf{q})} A(\textbf{p},\textbf{p}+\textbf{q})\\
&\approx -\frac{1}{(2\pi)^{2}}\int_{\varepsilon(\textbf{p})=0} \frac{ds(\textbf{p})}{|\partial \varepsilon/\partial \textbf{p}|} A(\textbf{p}-\textbf{q}/2,\textbf{p}+\textbf{q}/2).
\end{split}
\end{equation*}
The correction term to this relation is $O(|\textbf{q}|^{3})$ provided that $A(\textbf{p},\textbf{p}+\textbf{q})$ is sufficiently smooth.} The primary conclusion to draw from the weak-coupling analysis is that the pairing instability is in the spin-triplet channel. For a broad range of $\eta_{w}$, the dominant pairing solution is a ``$p_{y}$-wave state'' that is even in $k_{c}$ and odd in $k_{b}$. The associated pair wave function for $\eta_{w} = 1.0$ is shown in Fig. \ref{pairwave}.

Notice that there are more sign changes in the pairing gap than dictated by parities under $k_{c} \rightarrow -k_{c}$ or $k_{b} \rightarrow -k_{b}$. Most importantly, it has opposite signs between two nearly touching Fermi surface sheets. This fact is consistent with the observation in the previous section that the pair wave function has a propensity to change sign between two points $\hat{\textbf{k}}$ and $\hat{\textbf{k}}^{\prime}$ satisfying $\hat{\textbf{k}}+\hat{\textbf{k}}^{\prime}\approx \textbf{Q}$ (Fig. \ref{fs}).

Another notable feature is the gap suppression near $k_{c}=0$, as seen in Fig. \ref{pairwave}. A gap minimum exists on the outer Fermi surface at $k_{c}=0$ for the entire range of $\eta_{w}$ shown in Fig. \ref{pairing} and also on the inner Fermi surface for $\eta_{w} \lesssim 0.8$. For larger warping, two closely spaced accidental nodes develop around $k_{c}=0$ on the inner Fermi surface. The reason for this gap suppression can be again understood from the nesting condition $\hat{\textbf{k}}+\hat{\textbf{k}}^{\prime}\approx \textbf{Q}$. From Fig. \ref{fs}, one sees that most of crystal momentum pairs satisfying the nesting condition exist across the inner and outer Fermi surfaces. However, near $k_{c}=0$, the two Fermi surfaces almost coincide, and a pair of crystal momenta on the same Fermi surface, as well as a pair across different Fermi surfaces, can meet the nesting condition. Hence, there is frustration in the resulting sign change of the gap, leading to the suppression.

We remark that from the above consideration of the nesting properties, one may expect that that there exists a closely competing singlet solution, which has the same gap structure in the upper (or lower) half of the Brillouin zone as the dominant triplet solution but is even in both $k_{c}$ and $k_{b}$. However, the numerical result shows that this is not the case. Indeed, for solutions that are even in both $k_{c}$ and $k_{b}$, one has to additionally apply constraints given by Eq. (\ref{constraints}). As a result, what appears to be a nearly degenerate singlet solution is in fact significantly disfavored.

\section{Role of longer-range Coulomb interactions and spin-orbit coupling}
We also briefly considered the possible effects of off-site Coulomb interaction parameters introduced in Ref. \onlinecite{mckenzie2012prb}, and set them to be of order $U^2/t$ \cite{Raghu2012}.
However, if one neglects screening, there are a large number of these
parameters. A representative investigation of
the full parameter space would require another separate study.
Here, we just note that we did find that these interactions could
change the pairing symmetry and/or the number of accidental nodes in
the pairing amplitude. We also found that the repulsion between diagonal neighbors has a much more significant effect than that between nearest neighbors along or in the direction perpendicular to the zigzag chain.
Whether this sensitivity to longer0range interactions is an artifact of the weak-coupling analysis may only be revealed by numerical investigations of the intermediate- to strong-coupling regime.

Another important question is the direction of the $\vec{d}$ vector associated with the triplet pairing. This  will be determined by spin-orbit coupling and we plan to address this in a future study.

\section{Proposed future experiments}
One signature of triplet pairing is the temperature independence of the Knight shift in
nuclear magnetic resonance (NMR) below $T_c$ \cite{Maeno2012}. Hence, we propose measurements
using $^7$Li, $^{17}$O, and/or $^{95}$Mo NMR.
Recently, $^7$Li NMR measurements were performed on the normal state of this
material \cite{wu2014magnetic}.
In the past, $^{95}$Mo NMR has been difficult  due to the very low resonance frequency. However, recent advances in high-field NMR have seen its use in studies of a range of
molybdenum oxides \cite{Lapina}.
Evidence for the nodes in the superconducting
energy gap will be a power-law temperature dependence of thermodynamic quantities
such as the specific heat and the superfluid density at temperatures much less than $T_c$.
A recent theoretical study investigated the difference in the temperature dependence of the upper critical
field for triplet states in a quasi-one-dimensional superconductor
with and without nodes \cite{sepper2013}.
However, the differences are small and subtle, making definitive conclusions
about the presence or absence of nodes difficult.

\section{Conclusions}
We used asymptotically exact perturbative renormalization group methods
to analyze a minimal Hubbard model for Li$_{0.9}$Mo$_6$O$_{17}$.
We found that spin-triplet odd-parity superconductivity is the dominant instability,
with a gap having more sign changes than required by the point-group symmetry.
Hopefully, our study will stimulate further experimental studies of the superconducting
state of this material, to find definitive evidence for triplet pairing and/or
nodes in the energy gap.

\emph{Note added.} Recently we learned of a work by Lera and Alvarez \cite{lera15} that performed a multi-orbital RPA analysis of a similar Hubbard model and reached similar conclusions to ours.

\begin{acknowledgments}
This work was supported in part by the Office of Basic Energy Sciences, Materials Sciences and Engineering Division of the US Department of Energy under Contract No. AC02-76SF00515 (W.C., S.R.), and the Alfred P. Sloan Foundation (S.R.).
R.H.M. was supported in part by an Australian Research Council Discovery Project.
We thank D. Agterberg, S.-B. Chung, A. Jacko, and J. Merino for helpful discussions.
\end{acknowledgments}

\bibliography{triplet}

\end{document}